\documentclass[runningheads]{llncs}

\usepackage[T1]{fontenc}
\usepackage{graphicx}
\usepackage{listings}
\usepackage{booktabs} 
\usepackage{url}
\usepackage[inline,shortlabels]{enumitem}
\usepackage{tcolorbox}
\usepackage{todonotes}
\usepackage{wasysym}
\usepackage{changepage}
\usepackage{csquotes}
\usepackage{fmtcount}
\usepackage{xcolor}
\usepackage{tikz}
\usetikzlibrary{positioning,fit}
\usepackage[hidelinks]{hyperref}

\AtBeginDocument{}

\definecolor{OliveGreen}{cmyk}{0.64,0,0.95,0.40}

\newcommand{\Section}[1]{Section~\ref{#1}}
\newcommand{\Figure}[1]{Figure~\ref{#1}}
\newcommand{\Table}[1]{Table~\ref{#1}}

\newcommand{\HIDE}[1]{}
\newcommand{\TODO}[2][-]{{\color{red}\textbf{[TODO(#1): #2]}}}

\newcommand{\RQ}[1]{\textbf{RQ\textsubscript{#1}}}

\newenvironment{fancyquote*}[1][]{\def\attribution{#1}``\unskip\small\itshape\ignorespaces}{\unskip{''} \mbox{\textsc{\em\scriptsize\attribution}}}

\AtBeginDocument{%
  }

\title{Helpful but Fallible: Developer Experiences of AI Tools Under a Coordinated Industrial Roll-out}
\titlerunning{Helpful but Fallible: Developer Experiences of AI Tools\ldots}

\author{Andreas Bexell\inst{1,2}\orcidID{0009-0003-1356-303X} \and
Rushali Gupta\inst{1}\orcidID{0009-0006-3293-079X} \and
Lo {Gullstrand Heander}\inst{1}\orcidID{0000-0002-0695-4580} \and
Emma Söderberg\inst{1}\orcidID{0000-0001-7966-4560} \and
Per Runeson\inst{1}\orcidID{0000-0003-2795-4851} \and
Sigrid Eldh\inst{2,3,4}\orcidID{0000-0002-5070-9312} \and
Wei Shi\inst{5,2}\orcidID{0000-0002-4541-2098} \and
Konstantin Malysh\inst{1}\orcidID{0000-0002-3659-3093}}
\institute{Lund University, Lund, Sweden \and
Ericsson AB, Sweden \and
Mälardalen University, Västerås, Sweden \and
Carleton University, Ottawa, Canada \and
KTH Royal Institute of Technology, Stockholm, Sweden\\
\email{\{rushali.gupta, lo.gullstrand\_heander, emma.soderberg, per.runeson, konstantin.malysh\}@cs.lth.se}\\
\email{\{andreas.bexell, sigrid.eldh, wei.b.shi\}@ericsson.com}}

\authorrunning{A. Bexell et al.}

\begin{document}

\maketitle
 
\begin{abstract}
AI-enabled software development tools (AI-devtools) are being industrially adopted under strong expectations of productivity gains, yet developers' experiences of such roll-outs are underexplored. Organizations commit budgets, evaluate staff, and revise practice on a partial picture, since the evidence base is mainly tool evaluations, productivity metrics, and surveys, with few qualitative in-situ accounts of ongoing, coordinated roll-outs. We report a case study of a coordinated roll-out of AI-devtools at a large Swedish telecommunications company, investigating how developers experience the roll-out and how they anticipate their profession will change. We conducted semi-structured interviews with 12 software professionals across three sites, analyzed with process coding and thematic analysis, and interpreted through the extended Technology Acceptance Model (TAM2) as a post-hoc analytical lens. Our findings on use cases, productivity, frustrations, and tool limitations corroborate prior survey work. Beyond corroboration, the interviews surface a management--developer expectation gap that maps onto the TAM2 constructs of subjective norm and voluntariness, and show that participants weigh perceived \emph{risk} heavily, a factor that TAM2 and similar acceptance models do not represent. AI-devtools emerge as helpful but fallible assistants whose value is shaped by organizational expectations, system scale, and developers' skills.

\textbf{Data Availability:} \url{https://doi.org/10.5281/zenodo.20021475}

\keywords{AI coding assistants \and technology acceptance \and case study \and thematic analysis \and developer experience}

\end{abstract}

\section{Introduction}
\label{sec:introduction}

Across many industrial software organizations, AI-enabled software development tools (AI-devtools) are being rolled out as drivers of productivity and efficiency. Although the underlying models and assistants improve quickly, the experience of the practicing developers who adopt them under such coordinated roll-outs is uneven~\cite{pinto2024developer} and contested~\cite{chen2024impact}. The mismatch between expectation and experience could risk distorting performance appraisal, mis-allocating tool investment, weakening  individual skills (as routine reliance replaces practice) and the collegial interaction through which teams share knowledge, and leave organizations without a credible account of where the tools actually deliver value.
%
Existing empirical evidence on this transition comes predominantly from observational studies (e.g. Kumar~et~al.~\cite{kumar2025intuition}), tool evaluations (e.g. Pinto~et~al.~\cite{pinto2024developer}), and self-report surveys (e.g. Martinovi\'c and Rozi\'c~\cite{martinovic2025perceived}), reporting broad productivity gains in some settings, frustrations and reliability concerns in others, and largely neutral perceptions in still others. The findings are informative, and qualitative interview accounts of professional AI-assistant use are beginning to accumulate~\cite{chen2026beyond,klemmer_using_2024,mendes_youre_2024}; yet they mostly miss the experiences of a coordinated roll-out in a large production code base, including the social, organizational, and risk-laden reasoning that surrounds it. To our knowledge, in-depth accounts of such mandated, organization-wide roll-outs, studied in situ as they unfold, remain scarce.

In this paper, we present a case study of a coordinated AI-devtool roll-out at a large Swedish telecommunications company, drawing on semi-structured interviews with 12 software professionals at three sites, conducted approximately six months into the roll-out. The roll-out is itself a software process change, and understanding how developers experience it is a prerequisite for improving such adoption processes. The interviews provide an in-situ snapshot of the roll-out, capturing the participants' use cases, frustrations, expectations, and concerns.
Through thematic analysis, we address the following research questions:

\begin{description}
    \item[\RQ{1}] \emph{\small How do professional software developers experience a coordinated introduction of AI-devtools?}
    \item[\RQ{2}] \emph{\small How do developers anticipate their profession will change following the introduction of AI-devtools?} 
\end{description}

\begin{figure}[hbt]
\centering
\begin{tikzpicture}[
  scale=0.85, transform shape,
  font=\scriptsize,
  theme/.style={draw, rounded corners, align=left, text width=3.6cm, inner sep=2.5pt, fill=gray!8},
  construct/.style={draw, align=center, text width=2.8cm, inner sep=1.8pt, fill=white},
  risknode/.style={draw, dashed, thick, align=center, text width=2.8cm, inner sep=2.5pt, fill=gray!20},
  map/.style={->, >=stealth, gray!70!black},
  node distance=2.2mm and 18mm
]
\node[theme] (corp) {\textbf{Corporate Structure} (44)};
\node[theme, below=of corp] (rel) {\textbf{Relevance} (288)};
\node[theme, below=of rel] (exp) {\textbf{Experience} (94)};
\node[theme, below=of exp] (skills) {\textbf{Skills in Transformation} (24)};
\node[theme, below=of skills] (human) {\textbf{Human Aspect} (195)};
\node[theme, below=of human] (trust) {\textbf{Trust and Responsibility} (18)};
\node[construct, anchor=west] (sn) at (5.6, 0.15) {Subjective Norm};
\node[construct, below=0.9mm of sn] (vol) {Voluntariness};
\node[construct, below=0.9mm of vol] (img) {Image};
\node[construct, below=0.9mm of img] (jr) {Job Relevance};
\node[construct, below=0.9mm of jr] (oq) {Output Quality};
\node[construct, below=0.9mm of oq] (rd) {Result Demonstrability};
\node[construct, below=0.9mm of rd] (pu) {Perceived Usefulness};
\node[construct, below=0.9mm of pu] (moexp) {Experience};
\node[construct, below=0.9mm of moexp] (peou) {Perceived Ease of Use};
\node[draw, gray, inner sep=2mm, fit=(sn)(peou), label={[gray, font=\scriptsize]90:TAM2 constructs}] (tamframe) {};
\node[risknode, below=3mm of peou] (risk) {\textbf{Perceived Risk}\\ (not in TAM2)};
\draw[map] (corp.east) -- (sn.west);
\draw[map] (corp.east) -- (vol.west);
\draw[map] (corp.east) -- (img.west);
\draw[map] (rel.east) -- (jr.west);
\draw[map] (rel.east) -- (oq.west);
\draw[map] (rel.east) -- (rd.west);
\draw[map] (rel.east) -- (pu.west);
\draw[map] (exp.east) -- (moexp.west);
\draw[map] (exp.east) -- (peou.west);
\draw[map] (skills.east) -- (peou.west);
\draw[map] (corp.south east) to[out=-15, in=180] (risk.west);
\draw[map] (rel.south east) to[out=-15, in=178] (risk.west);
\draw[map] (exp.south east) to[out=-12, in=176] (risk.west);
\draw[map] (skills.east) to[out=-15, in=175] (risk.west);
\draw[map] (human.east) -- (risk.west);
\draw[map] (trust.east) -- (risk.west);
\end{tikzpicture}
\caption{The six themes with per-theme segment counts (left), mapped onto TAM2 constructs~\cite{venkatesh2000theoretical} (right) by authors 1--3 in two joint sessions (\Section{sec:analytical_lens}); subthemes are italicized in \Section{sec:results}. \textbf{Takeaway:} The themes populate all TAM2 constructs, with \emph{Image} the most thinly evidenced; reasoning about risk occurs in all six themes but maps onto no TAM2 construct, marking \emph{perceived risk} as a gap in the model.}
\label{fig:theme_map}
\end{figure}
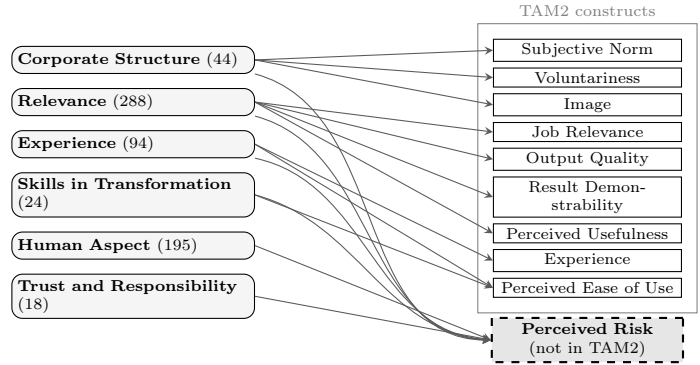

We interpret our findings through the extended Technology Acceptance Model (TAM2)~\cite{venkatesh2000theoretical} as a post-hoc analytical lens (\Figure{fig:theme_map}), and relate them to the EU's trustworthy-AI guidelines~\cite{european2019ethics}.

This study makes three contributions to the conversation on industrial adoption of AI-devtools.
\begin{enumerate*}[label={\em\protect\Ordinalstring{enumi},}]
    \item our findings on use cases, perceived productivity, frustrations, and tool limitations corroborate prior studies, the convergence between an in-situ account and survey evidence adding confidence to the emerging picture;
    \item the interviews surface a richer description of the tension with management than previously reported, namely pressure to adopt and demonstrate gains despite the tools' immaturity for large code bases, and mixed sentiments toward AI usage as an appraisal metric, connecting to the TAM2 constructs of \emph{Subjective Norm} and \emph{Image} yet sharpening them with concrete organizational mechanism;
    \item participants reason at length about \emph{perceived risk}: to their employment, of data leakage, deskilling, and losing the ability to understand and maintain the code base, a factor that shapes their adoption behavior yet is notably absent from TAM2 and subsequent acceptance models.
\end{enumerate*}

\section{Related Work}

AI-devtools are widely reported to boost developer productivity, yet accumulated evidence reveals a more contested picture. In a mixed-methods study of AI-assisted pair programming, Chen~\cite{chen2024impact} found \enquote{a positive impact on code quality and developer satisfaction}, but also reported challenges of lost autonomy, lacking trust, and questioned reliability of the generated code. Similarly, Pinto et al.~\cite{pinto2024developer} reported that the efficiency gains of Stackspot AI were unlocked only with specific knowledge of the tool, with frustration over inconsistency, history loss, and limitations on large code structures. On productivity, the respondents of Martinovi\'c and Rozi\'c~\cite{martinovic2025perceived} were largely neutral except for self-perceived programming efficiency, while Kumar et~al.~\cite{kumar2025intuition}, documenting in-house usage by 300 engineers over a year, reported a 31.8\% PR review cycle time reduction and 93\% of users wishing to continue.
%

Two interview studies come closest to ours. Chen~et~al.~\cite{chen2026beyond} complicated the productivity picture, finding AI tools beneficial for feature development and test creation but detrimental to maintenance, expertise retention, and feeling of ownership; Mendes~et~al.~\cite{mendes_youre_2024} described 14 developers' daily experiences, benefits, challenges, and coping strategies. Both focus on individual developers' experience of the assistants, rather than an organization-wide introduction of them.

The tools are also applied far beyond code completion. Coutinho et al.~\cite{coutinho2024role}, in a case study with 13 participants, and Sergeyuk et al.~\cite{sergeyuk2025using}, in a survey of 481 developers, together documented use for document and plan generation, ideation, feature implementation, test writing, triage, refactoring, and natural-language artifacts. The latter listed the main obstacles to organizational adoption as lack of need, inaccurate output, lack of trust, and lack of context understanding, obstacles mirrored in the frustrations of Pinto et al.~\cite{pinto2024developer} and Chen~\cite{chen2024impact}. Security-specific concerns were examined by Klemmer~et~al.~\cite{klemmer_using_2024} through 27 interviews complemented by 190 Reddit posts. Despite widespread concerns, their participants used AI assistants for security-critical tasks and the resulting mistrust led them to check AI suggestions much as they would human-written code.

Jensen et al. \cite{jensen2025managing} found that as the tools came into use in three organizations, expectations on code quality, productivity, and satisfaction largely persisted, whereas those on inter-team communication, project management, and business improvement faded as anticipated organizational benefits failed to materialize. Closest to our analytical lens, Shao and Ishengoma~\cite{shao2026empirical} analyzed acceptance through UTAUT extended with security concerns, finding that professional developers systematically refine generated code for maintainability and architectural alignment, and that social influence, rather than perceived usefulness or ease of use, was the strongest predictor of their adoption intention.

Prior work thus reports broad productivity gains alongside uneven and contested experiences, and qualitative interview accounts are emerging~\cite{chen2026beyond,klemmer_using_2024,mendes_youre_2024}. Existing studies examine individual adopters, or perceptions detached from the organizational process introducing the tools. Organizations therefore appraise staff and direct investment without an account of a \emph{coordinated, organization-wide roll-out} studied in situ as it unfolds. We address this with an interview-based case study of developers in an industrial setting where AI-enabled tools are introduced through a mandated, organization-wide process.

\section{Method}
\label{sec:method}

\subsection{Case Study Design}
\label{sec:case_study_design}

We designed the study as a single holistic case study, following Runeson and H{\"o}st~\cite{runeson2009guidelines}: the case is the coordinated roll-out of AI-devtools at a large Swedish telecommunications company, a highly specialized development organization with large closed-source code bases. Consistent with an \emph{interpretivist} stance, semi-structured interviews with software professionals at three sites are the primary source of evidence, since the phenomena of interest, developers' lived experience of the roll-out (\RQ{1}) and their anticipation of professional change (\RQ{2}), are accessible primarily through first-person accounts. This grounds the reported experiences in one organization's concrete roll-out at the price of statistical generalizability (\Section{sec:threats}). The interviews were transcribed and analyzed using thematic analysis~\cite{braun_using_2006,clarkeThematicAnalysis2017}, iterating over the six steps of Braun and Clarke: \emph{familiarizing yourself with your data, generating initial codes, searching for themes, reviewing themes, defining and naming themes,} and \emph{producing the report}.

\subsection{Case and Context}
\label{sec:case_context}

The study was conducted at the end of 2025 at three Swedish sites of the company, which develops software for telecommunications infrastructure. The adoption of AI tools was mandated by top management and not a bottom-up developer initiative. This mode of adoption likely influenced the perceptions and usage patterns we observed. The roll-out unfolded gradually over six months to a year, the latest version of the AI-devtools having been available for around six months at the time of the interviews. The mandated tool-chain combines a command line interface (CLI) agent, integrable into existing integrated development environments (IDEs), with an agentic development environment, and supports the Model Context Protocol (MCP) for loading source code and documentation into the model's context. Developer training was largely peer-to-peer, spreading through designated AI early-adopters within teams (see also \Section{sec:skills_in_transformation}), and integration of AI usage into performance appraisal was uneven. Some managers and units had adopted it as a goal metric, others had not, or not yet. The developers thus encountered the same mandated tool-chain under varying degrees of formalized adoption pressure.

\subsection{Data Collection}
\label{sec:data_collection}
\begin{table}[tb]
\centering
    \caption{Interview participants. \textbf{Takeaway:} Participants span 1.5–24 years of experience, providing variation across role and seniority within a single organizational context.}
    \label{tab:participants}
    \scriptsize
    \begin{tabular}{lllr@{\hspace{1.5em}}lllr}
        \toprule
          &  & \bf Job Title & \bf Exp. (sw/comp.) &
          &  & \bf Job Title & \bf Exp. (sw/comp.) \\
        \midrule
        P1 & \mars & Solution Tester & 6 y / 6 y & P7 & \mars & Software Developer & 1.5 y / 1.5 y \\
        P2 & \mars & Software Developer & 1.5 y / 1.5 y & P8 & \venus & Software Engineer & 15 y / 5 y \\
        P3 & \mars & Developer & 22 y / 20 y & P9 & \mars & Developer & 3 y / 3 y \\
        P4 & \venus & Software Developer & 2.5 y / 1.5 y & P10 & \mars & Software Researcher & 24 y / 12 y \\
        P5 & \mars & Software Designer & 18 y / 2 m & P11 & \mars & Developer & 1 y / 9 m \\
        P6 & \mars & Software Developer & 5 y / 5 y & P12 & \mars & Software Developer & 5 y / 2 m \\
        \bottomrule
    \end{tabular}
    \normalsize
\end{table}

We interviewed 12 software professionals (\Table{tab:participants}) from three company sites. Participants were recruited through a call, endorsed by management, that reached the developers with access to the AI-devtools; the twelve who volunteered constitute the full set of respondents. We judged this sample adequate since it spans three sites, roles from tester to researcher, and 1.5--24 years of experience, and later interviews raised few experiences not already present in earlier ones. The interviews were semi-structured, based on a protocol designed to take about 40 minutes (several ran over, the longest just over one hour), divided into
\begin{enumerate*}[label=\arabic*)]
    \item a biographic introduction to ease the participant in;
    \item a section on the individual AI-devtools the participant had used;
    \item a section on general experiences with the roll-out and tool-chain; and
    \item an exit catch-all question.
\end{enumerate*}
From the fifth interview onward, a question about the feedback mechanism for AI-devtools was added; participants interviewed before were not re-contacted, so we treat the smaller answer base for this topic as a limitation (\Section{sec:threats}). The complete protocol is in the replication package (\Section{sec:data_availability}).

The interviews were conducted by authors 1, 3, 7, and 8, in pairs, with overlapping interviewer constellations across sessions for continuity. To capture the roll-out in situ, interviews took place on site, during working hours, in the participants' usual work environment, while the roll-out was ongoing; two participants, ill on their interview occasions, joined over video call instead. Informed consent was secured in writing before the interviews, except for two individuals who consented online during the recorded session. Three interviews were conducted in Swedish, the rest in English.

\subsection{Data Processing}
\label{sec:data_processing}

The interviews were automatically transcribed and, when needed, translated into English (see \Section{sec:data_availability} for source code), then manually error-corrected against the original audio by the data-collection team joined by author 2. Through conducting, revising, and reading the transcripts, we became \emph{familiar with the data}. We then performed \emph{initial coding}~\cite{braun_using_2006,clarkeThematicAnalysis2017,charmaz_constructing_2014}, employing process coding, where every code starts with a gerund~\cite{saldana_2015} to capture the action and intention behind each segment. To calibrate, we coded the first interview in a group workshop; coding was then done in pairs and, as consensus grew, individually. This calibration by consensus, rather than independent double-coding, means that an inter-coder agreement coefficient is not defined for the majority of the material; we consequently report segment counts descriptively rather than as evidence of theme importance, since the \emph{keyness} of a theme does not depend on quantifiable measures~\cite{braun_using_2006}.

\subsection{Data Analysis}

Authors 1, 2, and 3, joined by author 4, iterated over the initial codes across several workshop sessions, creating, merging, renaming, and categorizing them in \emph{search for themes}, then \emph{reviewing} the emerging themes before a concluding workshop produced their final \emph{definitions and names}. The complete code book, with theme mappings, is in the replication package (\Section{sec:data_availability}). For example, the fragment ``when you start to scale it up to bigger system and projects, you really have to be careful because it can't really capture all the dependencies'' (P10, \Section{sec:relevance}) was assigned the process codes \emph{describing AI limitations} and \emph{reflecting on AI user experience}, and organized under themes \emph{Experience} and \emph{Relevance}, contributing to the latter's \emph{scale limits} subtheme; a segment may thus belong to more than one theme.

\subsection{Analytical Lens}
\label{sec:analytical_lens}

During interpretation, we adopted TAM2~\cite{venkatesh2000theoretical} as a post-hoc analytical lens to organize our discussion. The study was designed to be exploratory and inductive and imposing an acceptance model during interviewing or coding would have narrowed the account to the model's constructs. TAM2 was instead selected after theme construction was complete, when the management--developer tension in the data called for a vocabulary from acceptance theory. We chose TAM2 over alternatives because it explicitly models \emph{Subjective Norm}, \emph{Voluntariness}, and \emph{Image}, the constructs a mandated roll-out activates. The original TAM lacks social influence altogether, whereas UTAUT and UTAUT2 fold it into a single construct~\cite{venkatesh2003user,venkatesh2012consumer}, blurring the distinction between normative pressure and formal mandate that our data exhibits. Social influence dominating adoption intention in a recent UTAUT study of software professionals~\cite{shao2026empirical} further motivates a lens that decomposes it.

Authors 1, 2, and 3 performed the mapping over two sessions, discussing disagreements until interpretative convergence; the original codes were not re-coded, so the mapping operates at the level of themes and subthemes (\Figure{fig:theme_map}), and is used descriptively rather than predictively. Two boundary observations follow. First, material on \emph{perceived risk} did not map onto any TAM2 construct; rather than force it, we report it as a gap in the model (\Section{sec:risk}). Second, all constructs received some support, but \emph{Image} is the most thinly evidenced, supported mainly by segments on the legitimization of AI use (\Section{sec:corporate_structure}). We acknowledge the post-hoc selection as a threat to construct validity in \Section{sec:threats}.

\section{Results}
\label{sec:results}

Thematic analysis revealed six themes: Corporate Structure, Experience, Human Aspect, Relevance, Skills in Transformation, and Trust and Responsibility. Our unit of analysis is the \emph{coded segment}: a contiguous transcript excerpt assigned at least one process code. \Figure{fig:theme_map} reports the per-theme counts, in which each segment is counted once per theme it belongs to, so the counts sum to 663 theme assignments over a smaller set of segments; Relevance (288) and Human Aspect (195) carry 73\% of those assignments.

\subsection{Corporate Structure}
\label{sec:corporate_structure}

This theme captures how the hierarchical organization imposes norms and expectations that would not be present in a non-professional setting. There is a clear \emph{normative pressure}, communicated as an expectation that everyone will ``leverage AI, accelerate your own work'' (P11) and increasing further up the chain-of-command. This creates a \emph{management--developer gap}: managers have access to a separate set of AI tools focused on management productivity, not the developers' tool-chain, so they promote AI-devtools without first-hand experience or situated understanding of what may make them under-perform. \begin{fancyquote*}[P2]we get a pressure on us that the AI should solve everything. But we know it’s not that good. It creates a huge gap between the reality you’re in as a software developer and the reality that is being sold to [the managers].\end{fancyquote*}

Attitudes to \emph{legitimization of AI use} are mixed and participants are largely neutral about having AI usage included in performance appraisal. Some find it strange to be evaluated on it, others see it as rewarding an interest in AI, or welcome the endorsement of a previously dubious way of working: \begin{fancyquote*}[P12]it felt like cheating using AI. It was like a little bit of rogue-like behavior, but now it’s very different. It’s encouraged. So, the stigma of using AI is disappearing.\end{fancyquote*}

\subsection{Experience}
\label{sec:experience}

On \emph{non-determinism}, participants report mixed experiences of working with a tool of unclear capabilities: \begin{fancyquote*}[P1]Sometimes it’s surprisingly good and sometimes it’s surprising that it can’t solve it.\end{fancyquote*} Despite this, they generally find the tools useful and supportive (see also \Section{sec:relevance}), especially when integrated into their existing work environment, eliminating the friction of copying context between interfaces. They describe an \emph{iterative partnership}: the developer describes the problem, the tool generates code, the developer reviews, checks, and amends it, and instructs the tool to revise, treating it as an assistant with limited capacities: \begin{fancyquote*}[P1]I see it as a slightly dim colleague who is very good at doing the boring stuff.\end{fancyquote*} Two participants note that as prompts grow more detailed, the perceived benefit diminishes: \begin{fancyquote*}[P2]You have to have such a good prompt so that [you] solve the task yourself. [...] And write down in the prompt exactly what it needs to do.\end{fancyquote*}

The partnership is fragile. \emph{Errors cascade}: once the model goes in the wrong direction, it is hard to course correct, and the output spirals into what P2 calls ``a vicious circle of errors''. In some workflows the tools may generate ``a completely new microservice'' (P2) whose workings developers do not understand. \emph{Context limitations} compound this when large systems strain the limited context windows of the LLMs, which may hang, crash, or forget instructions as the window fills. Some participants alleviate this with Model Context Protocol (MCP) integrations, vector databases indexing the source code, and structured rule files, becoming context engineers as well as software engineers.

\subsection{Human Aspect}
\label{sec:human_aspect}

The \emph{Human Aspect} theme concerns how AI-devtools affect human interaction and the developer's role: how work, roles, and work environments have changed, and how participants expect the profession to evolve. On the \emph{changing role}, there is a consensus that the tools make some tasks substantially faster, shifting the focus of development to the point where the developer's role changes, and automating some tasks so far that certain roles become redundant: \begin{fancyquote*}[P11]I think that if you’re a front-end developer, those jobs are already going away.\end{fancyquote*} Participant 12 describes one case of leading a project where they encouraged junior developers to rely on AI, where the participant's own work became guiding and reviewing solutions rather than writing them: \begin{fancyquote*}[P12]I encouraged him to use AI. It really didn’t matter which solution he found out. [...] I didn’t erase his solutions as long as they fulfilled everything. And we had a lot of communication because at that time, the AI solutions back then, they could easily loop. [...] And then you had to find different ways to approach that.\end{fancyquote*}

A chat-based AI-devtool can also \emph{reduce collegial interaction}: this cuts interruptions, but less human interaction can erode a sense of workplace community, and two participants worry that some of the social glue may be lost: \begin{fancyquote*}[P10]If you sit with a colleague, you have much more kind of talk about other things, take a coffee, this kind of things. [...] now everyone is just sitting with their own AI tools and then you don’t really talk to each other as always. So it’s a good technical thing, but it’s not a social thing.\end{fancyquote*} Finally, there is \emph{replacement fear}: participants fear the tools may disrupt the labor market for some roles, share this fear with colleagues, and do not miss the irony that developers may be made redundant by software, ``making ourselves redundant [...] I think we're destroying ourselves'' (P3).

\subsection{Relevance}
\label{sec:relevance}

The theme \emph{Relevance} captures participants' sentiments around the perceived relevance and usefulness of the introduced AI tools: use cases, relevance to each participant's context and tasks, frequency of use, and tool limitations.

\paragraph{Productivity.} Participants express that the AI-devtools increases their productivity when used for languages the agents are trained in: \begin{fancyquote*}[P11]this autumn I discovered the coding agents. You can’t use them for everything [...] but I’m working only in Python at the moment, so they are very proficient in that, just helping me become much more efficient.\end{fancyquote*}

\paragraph{Scale limits.} Several participants lift breakdown on large real-world projects as the primary limit to the tools' relevance, and this is the most-mentioned single concern in the theme: \begin{fancyquote*}[P10]
    for smaller tasks, like the one I showed here, containing just maybe three or four files, [...] it’s all excellent [...] when you start to scale it up to bigger system and projects, you really have to be careful because it can’t really capture all the dependencies and all the things [...] If you’re not careful, it will kind of break this very, very easily in some way.
\end{fancyquote*}

We place this account under \emph{Relevance} rather than \emph{Trust and Responsibility}: the point is that \emph{usefulness} is bounded by system scale, and the caution the tools demand is a consequence of that bound. Some participants are skeptical the tools will ever handle their part of the industry; working with specialized software for specialized hardware, in specialized languages, P5 (who designs proprietary hardware) sees no role for them until a model is trained on the relevant design-level specifications.

\paragraph{Use cases.} A wide variety of daily tasks are mentioned: writing code (all interviews), unit tests (9), information search (9), documentation (7), and debugging (5), with almost all interviews surfacing a use case unique to that participant. Two stand out. Generating documentation is a gateway that convinced participants to adopt AI tools seriously: \begin{fancyquote*}[P1]No one wants to write documentation. So when you show that [...] you can just ask the AI to use the template and generate documentation. Just as good or even better than what you would have made the effort to do yourself.\end{fancyquote*} Test case generation is a dramatic time saver for repetitive structures (notably, the one participant who uses the tools for refactoring feels slower for it). The generated tests are not always trustworthy, however; iterating on a failing test, the tool may silently remove the obstacle rather than solve the problem: \begin{fancyquote*}[P2]after a while it just says: ‘It’s too hard to set up, I’ll just remove the test case instead’. Or ‘I comment out this assertion and now just pass the test’.\end{fancyquote*}

\paragraph{Expectations.} Participants expect the tools to support more of the development process in the future, especially tedious tasks such as test creation and documentation, but also to enhance tasks in novel ways rather than automate them, e.g., finding bug report duplicates or searching large documentation sets, bringing gains in productivity, code quality, and adherence to sound design principles.

\subsection{Trust and Responsibility}
\label{sec:trust_responsibility}

This theme, the smallest, concerns trust in and reliance on the AI-devtools, and who is responsible when AI-generated code is wrong. Participants agree that \emph{human accountability} is non-negotiable: human oversight is required, and the human engineer is ultimately responsible and \begin{fancyquote*}[P1] cannot blame the AI\end{fancyquote*} when production crashes. Even when using AI, as P4 puts it, ``we as a human are still responsible of what we are submitting.'' That responsibility is sharpened by a \emph{calibrated distrust} where some tools are trained to be positive and agreeable, calling every suggestion an ``excellent idea'' (P10), which makes them difficult to trust. Several participants relate asking the AI to fix an issue only to find the tests still failing, concluding that its results cannot be trusted without verification.

\subsection{Skills in Transformation}
\label{sec:skills_in_transformation}

This theme concerns how AI knowledge diffuses through the organization, what skills might erode, and what new skills practitioners need. On \emph{knowledge diffusion}, several participants feel the organization lacks a mature, systematic structure for AI skill development, leaving knowledge transfer to motivated individuals who organize informal presentations and workshops; one such organizer notes that colleagues often do not know which tools are available, and calls for someone to drive the topic \begin{fancyquote*}[P9]because these things are happening so fast\end{fancyquote*}.

On \emph{deskilling}, participants worry that when reliance on AI becomes routine, engineers forget skills they previously possessed: \begin{fancyquote*}[P12]Everyone I've talked with in the previous company, they always felt like they forgot coding when they used AI. Stuff that is so basic that everyone knew it, they lost it.\end{fancyquote*} The worry extends to a future generation who may never acquire the understanding needed to audit AI-generated code, and to retaining code-base-specific knowledge, much of it implicit and unwritten, as the system is increasingly built by machines. Paradoxically, as more generated code must be scrutinized and fitted together, developers may need more, not less, skill to keep up.

Developing the \emph{new skills} this demands, such as precise prompting, critical evaluation of output, and context engineering (as also reported by Pinto et~al.~\cite{pinto2024developer}), benefits from prior software development knowledge: \begin{fancyquote*}[P3]If you are a developer, you can prompt the AI much better than if you are not a developer. Otherwise, you’re more likely to let AI dictate the terms.\end{fancyquote*} At the same time, \emph{onboarding} of junior developers goes faster, the tools guiding newcomers through the code base while accelerating their coding skills; combined with senior coaching, participants relate, this pulls new members forward faster than before. Overall, participants agree that developers with specific product knowledge and skill in working with AI tools will be more valuable to the company in the future.

\section{Discussion}
\label{sec:discussion}

For previously studied areas, our results corroborate prior work. Participants report increased productivity~\cite{kumar2025intuition,martinovic2025perceived}, a wide range of use cases~\cite{coutinho2024role,sergeyuk2025using}, and frustrations~\cite{chen2024impact,pinto2024developer} similar to those reported before, and their verification-heavy way of working matches the systematic refinement of generated code reported by Shao and Ishengoma~\cite{shao2026empirical}, though tied here to explicit reasoning about risk. Beyond corroboration, our case adds two elements: the mandated roll-out context, with appraisal-linked adoption pressure and a management--developer expectation gradient in which anticipated gains grow with hierarchical distance from the code; and developer-articulated \emph{risk} as a factor shaping adoption yet absent from the acceptance models commonly applied to AI-devtools. The convergence with prior findings suggests these experiences are not idiosyncratic to the studied organization. We make no statistical claim, but the alignment supports transferring the themes as analytic categories to comparable settings, i.e., large, specialized code bases under coordinated roll-outs.

Concerning \RQ{1} (experience of the roll-out), our participants gain productivity and find the tools useful for a variety of tasks, but also limited, demanding of specialized knowledge, and requiring oversight to produce reliable artifacts. We further find a management--developer tension, stemming from developers' apprehension that they and the tools cannot deliver the gains managers expect, and a continuous evaluation of risk when deciding how to interact with the tools and apply the results. Concerning \RQ{2} (anticipated change), participants expect their role to shift from programming to review and oversight, domain knowledge to outweigh general software engineering skills, and, for some, future advances to make human developers redundant.

In the following, we apply TAM2 (\Section{sec:tam}) to understand the tension between managers' expectations and developers' experiences (\Section{sec:corporate_structure}), and elaborate on alignment, risk, and trust, which TAM2 does not capture, in \Section{sec:risk}.

\subsection{Technology Acceptance}
\label{sec:tam}

Simplified, TAM2~\cite{venkatesh2000theoretical} models \emph{Subjective Norm} (possibly offset by \emph{Experience} and \emph{Voluntariness}), \emph{Image, Job Relevance, Output Quality} and \emph{Result Demonstrability} to affect \emph{Perceived Usefulness}, which together with \emph{Perceived Ease of Use} affects \emph{Intention to Use} and \emph{Usage Behaviour}. We relate our themes to these constructs as mapped in \Figure{fig:theme_map}.

The strongest signal in our data is \emph{Subjective Norm}, where the corporate structure forms a normative pressure to adopt, communicated both explicitly and through individual appraisal tied to compensation. Shao and Ishengoma~\cite{shao2026empirical} likewise found social influence the strongest predictor of adoption intention among software professionals ($\beta = 1.065$, $p < 0.001$), where earlier acceptance research emphasized perceived usefulness and ease of use; our account expands that result by describing the mechanism through which the pressure is exerted. This creates friction, as many participants judge the anticipated productivity increase inflated, and note that anticipated benefits grow with management level. That managers advocate a technology they do not themselves use is a recognized pattern in the diffusion of innovations~\cite{rogers2003diffusion}; specific here is the coupling of that advocacy to appraisal, and the expectation gradient along the hierarchy. The related construct of \emph{Image} is more thinly evidenced, supported mainly by the legitimization subtheme (\Section{sec:corporate_structure}), where the roll-out turned what felt like ``cheating'' into endorsed behavior.

On the usefulness side, participants find the tools relevant to their tasks (\emph{Job Relevance}), of context-dependent \emph{Output Quality} (pleasing for documentation, yet elsewhere failing local design rules), and productive in ways they sometimes feel unable to demonstrate to the extent expected of them (\emph{Result Demonstrability}). \emph{Perceived Usefulness} is thus generally positive but contested by limited applicability to specialized products and large legacy code bases in uncommon languages, and \emph{Perceived Ease of Use}, described as ``convenient'', can turn to disappointment against a steep learning curve with little guidance about the tools' limitations.

Although TAM2 captures normative pressure, ease of use, and perceived usefulness, it does not represent \emph{perceived risk} as a construct that affects adoption; nor, in their published form, do related frameworks such as the Unified Theory of Acceptance and Use of Technology (UTAUT)~\cite{venkatesh2003user} and UTAUT2~\cite{venkatesh2012consumer}. Yet many of our participants assess risk continuously, and that assessment governs how they interact with the tools and the generated artifacts (\Section{sec:trust_responsibility}).

\subsection{AI Alignment and Risk}
\label{sec:risk}

Reasoning about risk was not confined to the Trust and Responsibility theme (18 segments). Across the corpus, 22 distinct segments carry an explicit risk code, spanning all six themes, so we treat \emph{perceived risk} as a cross-cutting concern. To organize the risks, we relate them to two established scaffolds, the Ethics Guidelines for Trustworthy AI from the EU High-Level Expert Group~\cite{european2019ethics} and the Domain Taxonomy of AI Risks by Slattery et al.~\cite{slattery2024ai}, both of which have guided related discussions of misalignment risks in industrial software engineering~\cite{gupta2025ai}.

Five of the participant risks map onto these frameworks: \emph{data leakage} of prompts, source code, and proprietary documents; \emph{misalignment}, where generated code drifts from local rules or cascades errors, mitigated by keeping a human in the loop; \emph{deskilling and loss of code-base understanding}, compounding over time in long-lived systems; \emph{job loss}, falling hardest on juniors; and \emph{compromised models} injecting harmful code, against which current review practices were not designed. Participants frame these more concretely and personally than the population-level frameworks do. A sixth maps onto neither: the concern that anticipated productivity gains may not justify the investment in tooling, licensing, and infrastructure (\Section{sec:corporate_structure}). A single observation is not sufficient to argue for extending risk frameworks, but, taken together, our participants weigh perceived risk continuously when deciding how, when, and on which tasks to use AI tooling. We believe this points to a candidate direction for future acceptance models in AI-assisted software engineering: representing \emph{perceived risk} as an additional construct affecting adoption. Shao and Ishengoma~\cite{shao2026empirical} moved in this direction by adding \emph{security concerns} as a determinant to UTAUT, signalling the role of trust and perception of risk in adoption. Our participants' risks extend beyond security to deskilling, job loss, and the return on the investment itself, which suggests the broader construct of perceived risk rather than security alone.

Several opportunities remain for future work. Our interviews probed developers' \emph{experience} of the roll-out, not the coordination mechanisms themselves; studying those mechanics, and the managers' side of the expectation gap, is a natural next step. Broadening participation to other organizations, and adding longitudinal and mixed-method designs, could show whether the themes recur and relate perceived gains to actual outcomes.
\section{Threats to Validity}
\label{sec:threats}

Consistent with our interpretivist stance, we discuss threats to validity using the terminology suggested by Guba~\cite{guba1981criteria}.

\textbf{Credibility.}
Answers may be influenced by question framing, participants' relationship to the company, and the organizational narrative around AI-devtools. To reduce leading effects, we used open-ended, neutrally phrased questions, moving from biography and concrete tool use to sensitive topics such as performance management, with interviewers trained together. Self-reported data is also subject to recall and social desirability bias; the in-situ design (\Section{sec:data_collection}) aids recall, and we emphasized anonymity and the absence of managerial access to raw data to encourage candor. Furthermore, our design captures the developer perspective only; managers' accounts were not collected, so the reported management--developer misalignment is one-sided testimony about a two-sided relationship.

\textbf{Confirmability.}
All authors are familiar with software engineering and the discourse around AI-devtools, which risks shaping our interpretation. We therefore based our analysis on verbatim transcripts rather than notes, and report quotes illustrating both enthusiastic and skeptical perspectives, avoiding treating any single quote as representative without corroboration. A specific threat is our post-hoc use of TAM2, which risks retrofitting data to the model; we mitigated this by (i) documenting the mapping procedure (\Section{sec:analytical_lens}), (ii) using TAM2 descriptively rather than predictively, and (iii) noting where participants' reasoning, particularly around risk, does not map cleanly onto it. Readers should treat the TAM2-based discussion as one interpretation rather than the only valid framing.

\textbf{Dependability.}
To make the process transparent, we maintained a documented protocol, overlapped interviewers across sessions, and used a multi-stage coding process: a calibration workshop on the first interview, then pair and individual coding with overlapping constellations, and iterative theme refinement. We report no inter-coder agreement coefficient, since the consensus-based calibration described in \Section{sec:data_processing} does not yield one. One question was added from the fifth interview onward, without re-contacting earlier participants, so its answer base is smaller. We provide the protocol, code book, and segment locations in our replication package; for privacy reasons we cannot release full transcripts, which limits full replicability, but we aim to make the analytic steps traceable enough to evaluate.

\textbf{Transferability.}
Our study focuses on 12 professionals in a single large telecommunications company, in one country, during an early phase of a coordinated roll-out. Voluntary participation may favor individuals with stronger opinions; although we observe a diversity of views, we cannot rule out self-selection bias, nor claim statistical representativeness. As is common in qualitative research, our findings should be read as thematic insights from a particular context, and where they align with prior work they offer mutual corroboration; where they diverge, they may point to context-specific factors such as organizational structure or local AI governance. We therefore encourage caution in extrapolating beyond settings that resemble ours.

\section{Conclusions}
\label{sec:conclusions}

Participants describe AI-devtools as a helpful but fallible assistant whose limitations become pronounced in large, specialized code bases. Under the mandated roll-out, anticipated gains grow with hierarchical distance from the code: managers advocate tools they do not use in the developer tool-chain themselves, and developers report a gap between the productivity promised to them and what they observe. Interpreted through TAM2, this normative pressure meets context-dependent assessments of value, where the narrative of becoming an ``AI company'' sometimes conflicts with what developers can demonstrate. Participants also weigh risks continuously, from data leakage and misalignment to deskilling, loss of code-base understanding, and job security; these are largely covered by existing AI-risk frameworks, yet risk remains absent from TAM2, UTAUT, and UTAUT2 in their published form. We argue that \emph{perceived risk} should be treated as a first-class construct in future models of AI-tool acceptance. For practice, the accounts of our 12 participants suggest that roll-outs in comparable settings could benefit from calibrating expectations, investing in training and knowledge sharing, and keeping humans meaningfully in the loop with clear responsibility boundaries for AI-generated artifacts.

\section{Data Availability}
\label{sec:data_availability}

Our replication package contains the interview protocol, the code book with segment locations and tags (sufficient to re-create the paper's counts and figures), and the transcription and translation scripts; for privacy, full transcripts are not included. Available at \url{https://doi.org/10.5281/zenodo.20021475}.

\begin{credits}
\subsubsection{\ackname} We thank the industry participants for openly sharing their experiences. This work was partially supported by the Wallenberg AI, Autonomous Systems and Software Program (WASP) funded by the Knut and Alice Wallenberg Foundation, and the Competence Centre NextG2Com funded by the VINNOVA program for Advanced Digitalisation (grant 2023-00541).
\subsubsection{\discintname} The authors employed by Ericsson AB report their affiliation; the authors declare no other competing interests.
\end{credits}

\bibliographystyle{splncs04}
\bibliography{refs}

\end{document}